\documentclass[aps,prc,preprint,superscriptaddress]{revtex4-1}

\usepackage{graphics}

\begin{document}


\title{$^{50}$Ti($d$,$p$)$^{51}$Ti, single neutron energies in the $N=29$ isotones and the $N=32$ subshell closure}



\author{L. A. Riley}
\affiliation{Department of Physics and Astronomy, Ursinus College,
  Collegeville, PA 19426, USA}

\author{J. M. Nebel-Crosson} \affiliation{Department of Physics and Astronomy, Ursinus College,
  Collegeville, PA 19426, USA}

\author{K. T. Macon} \affiliation{Department of Physics and Astronomy, 
Louisiana State University, Baton Rouge, Louisiana 70803, USA}

\author{G. W. McCann} \affiliation{Department of Physics, Florida
  State University, Tallahassee, FL 32306, USA}

\author{L. T. Baby} \affiliation{Department of Physics, Florida
  State University, Tallahassee, FL 32306, USA}

\author{D. Caussyn} \affiliation{Department of Physics, Florida
  State University, Tallahassee, FL 32306, USA}

\author{P. D. Cottle} \affiliation{Department of Physics, Florida
  State University, Tallahassee, FL 32306, USA}

\author{J. Esparza} \affiliation{Department of Physics, Florida
  State University, Tallahassee, FL 32306, USA}

\author{K. Hanselman} \affiliation{Department of Physics, Florida
  State University, Tallahassee, FL 32306, USA}

\author{K. W. Kemper} \affiliation{Department of Physics, Florida
  State University, Tallahassee, FL 32306, USA}

\author{E. Temanson} \affiliation{Department of Physics, Florida
  State University, Tallahassee, FL 32306, USA}

\author{I. Wiedenho\"ver} \affiliation{Department of Physics, Florida
  State University, Tallahassee, FL 32306, USA}

\date{\today}

\begin{abstract}

A measurement of the $^{50}$Ti($d$,$p$)$^{51}$Ti reaction at 16 MeV was performed
using a Super Enge Split Pole Spectrograph to measure the magnitude of the $N=32$ subshell gap in Ti.  Seven states were observed that had 
not been observed in previous ($d$,$p$) measurements, and the \textit{L} transfer values 
for six previously measured states were either changed or measured for the first time.
The results were used to determine single neutron energies for 
the $p_{3/2}$, $p_{1/2}$ and $f_{5/2}$ orbitals.   
The resulting single neutron energies in $^{51}$Ti confirm the existence of the $N=32$ gap in Ti.  These single neutron energies and those from previous measurements 
in $^{49}$Ca, $^{53}$Cr and $^{55}$Fe are compared to values from a covariant 
density functional theory calculation.   
\end{abstract}


\maketitle


\section{Introduction}

The structure of atomic nuclei is strongly influenced by shell effects.  The most prominent examples of such shell effects are the nuclei with closed major shells of protons or neutrons – or both.  Such nuclei have spherical structure, indicated through high $2_1^+$ state energies, small values of 
$B(E2; 0_{gs}^+ \rightarrow 2_1^+)$ and characteristic signatures in the systematic behavior of nuclear masses.

Within major shells, subshell closures occur when there are sufficiently large energy gaps between orbits.  While the effects of such subshell closures are less pronounced than those of major shell closures, they can still be seen in $2_1^+$ state energies, $B(E2; 0_{gs}^+ \rightarrow 2_1^+)$ values and nuclear mass systematics.

However, the most direct way to infer a subshell closure is through a deduction of single particle energies using a single nucleon transfer reaction on a semi-magic target.  For example, the $^{48}$Ca($d,p$)$^{49}$Ca reaction \cite{Uo94} shows that there is a gap of 1.7 MeV between the lowest neutron orbit in the \textit{fp} shell, $p_{3/2}$, and the next lowest orbit, which is its spin-orbit partner $p_{1/2}$.  There is another gap of 1.7 MeV between the $p_{1/2}$ orbit and the next orbit, which is $f_{5/2}$.  Single particle orbits generally fragment into a number of states, and a sensitive $(d,p)$ measurement detects all of the significant fragments so that the single particle energy for a particular orbit can be determined as the centroid of the observed strength.

\begin{figure}[h]
  \begin{center}
    \scalebox{0.33}{
      \includegraphics{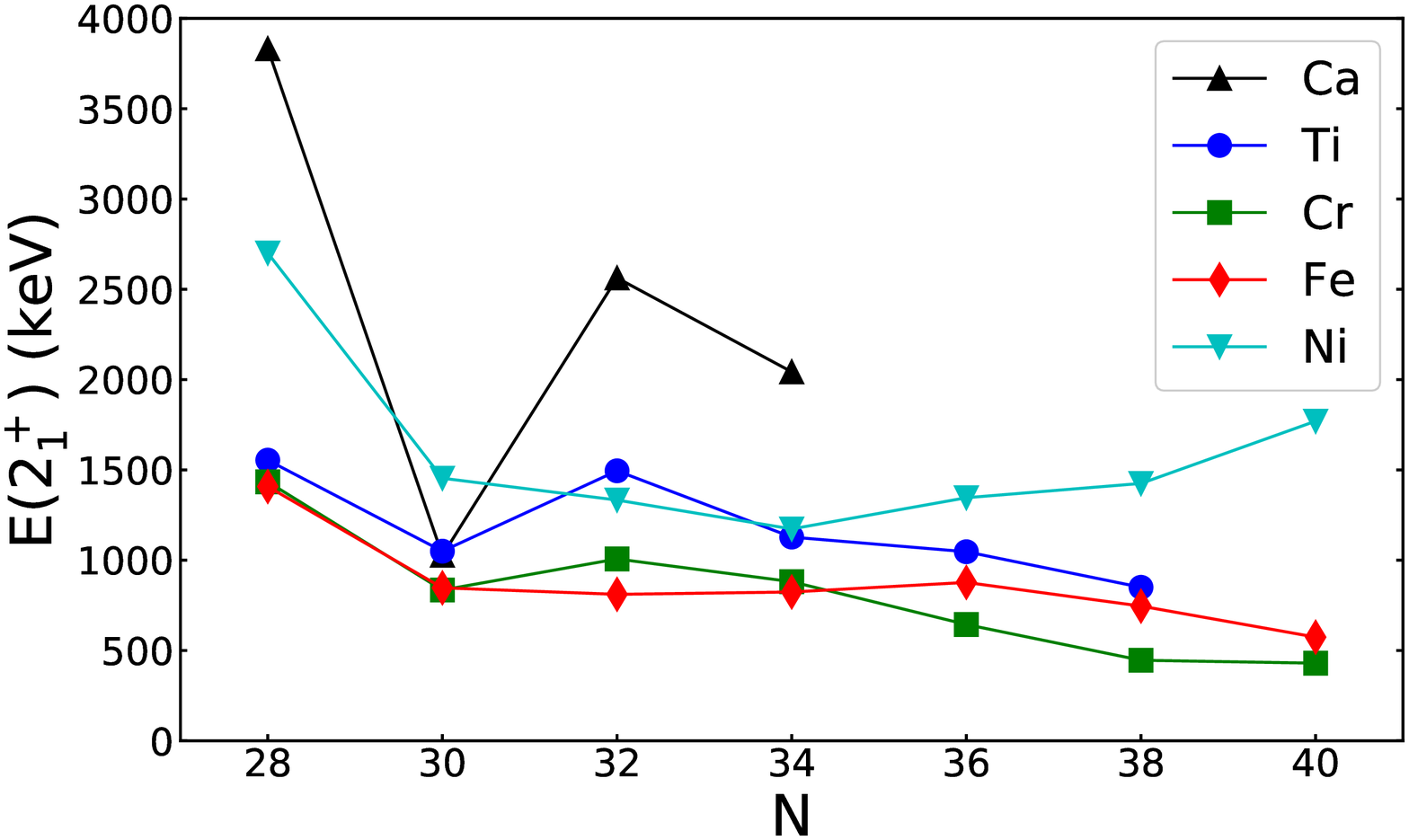}
    }
    \caption{\label{fig:e2} (Color online) $E(2_1^+)$ for the $N=28-40$ isotopes of Ca, Ti, Cr, Fe and Ni. \cite{NNDC48,NNDC50,NNDC52,NNDC54,NNDC56,NNDC58,NNDC60,NNDC62,NNDC64,NNDC66,NNDC68}}
  \end{center}
\end{figure}

The $N=32$ subshell gap between the $p_{3/2}$ and $p_{1/2}$ orbits in the neutron-rich isotopes of Ca and Ti provides an excellent example of the behavior of nuclei in the neighborhood of a subshell gap.  Figure~\ref{fig:e2} shows the systematic behavior of the energies of the lowest $2_1^+$ states, 
$E(2_1^+)$, in the even-even $N \geq 28$ isotopes of Ca, Ti, Cr and Fe.  While the largest values of $E(2_1^+)$ occur for all four elements at the $N=28$ major shell closure, the $E(2_1^+)$ values peak again at $N=32$ in Ca and Ti, but not in Cr and Fe.  We can infer from this behavior that the $N=32$ subshell gap exists in Ca and Ti, but narrows in Cr and Fe.  A recent mass measurement of the neutron-rich Ti isotopes \cite{Le18} provided the same conclusion - that the $N=32$ subshell closure exists in the Ca and Ti isotopes but not in the Cr isotopes.

If the N=32 subshell gap closes in the transition from Ti to Cr, as it appears to do, then it must be because the energy of the $f_{5/2}$ neutron orbit is higher than that of the $p_{1/2}$ orbit in Ti but decreases significantly in Cr.  After all, the spin-orbit splitting of the $p_{3/2}$ and $p_{1/2}$ neutron orbits is not likely to narrow significantly in the transition from Ti to Cr.  The extant study of the $^{50}$Ti($d,p$)$^{51}$Ti reaction \cite{Bar64} has the $f_{5/2}$ strength concentrated in two states of comparable (and large) strength that are 3 MeV apart.  One of those states exists at an excitation energy - 5139 keV - at which there is a relatively high density of states and where we would not generally expect such a large concentration of strength in a single state.

Here we present a new measurement of the $^{50}$Ti($d,p$)$^{51}$Ti reaction in which we changed or determined for the first time angular momentum (\textit{L}) transfer values on six previously known states and observed seven states that had not been observed in the
previous ($d$,$p$) measurement (\textit{L} was determined for two of these new states).  In particular, we observed a significant amount of higher-lying $L=3$ strength distributed among several states.  We did not observe the single strong $L=3$ state at 5139 keV reported in Ref. \cite{Bar64}.  In addition, we compare the single neutron energies we extract from the present results on $^{51}$Ti and previous results from $^{49}$Ca, $^{53}$Cr and $^{55}$Fe to a calculation of single neutron energies using covariant density functional theory. Our results suggest that a new measurement of the $^{49}$Ti($t$,$p$)$^{51}$Ti reaction should be performed to clarify the $f_{5/2}$ single neutron energy in $^{51}$Ti.  Furthermore, the $^{54}$Fe($d$,$p$)$^{55}$Fe reaction should 
be remeasured to investigate an anomalous result - the collapse of the $p_{3/2}-p_{1/2}$ spin-orbit splitting - that appears in the extant results on this reaction.

\section{Experimental details}

A deuteron beam, produced by a SNICS (Source of Negative Ions by Cesium Sputtering) source with a deuterated titanium cone, was accelerated to an energy of 16 MeV by the 9 MV Super FN Tandem Van de Graaff Accelerator at the John D. Fox Laboratory at Florida State University. The beam was delivered to a 0.45~mg/cm$^{2}$ Ti target enriched to $90\%$ in \textsuperscript{50}Ti that was mounted in the target chamber of the Super Enge Split-Pole Spectrograph. The spectrograph was rotated from scattering angles of 10$^\circ$ to 50$^\circ$ at increments of 5$^\circ$ to capture angular distributions of protons from the  \textsuperscript{50}Ti($d,p$)\textsuperscript{51}Ti reaction. Protons from the reaction were guided by magnetic fields to the focal-plane detector consisting of an isobutane-filled ion chamber with two proportional-counter anode wires at positive potential running the length of the detector above a Frisch-grid. The Frisch-grid allowed for cleaner timing and spatial resolution. The anode signals measured the charge each wire collected from the upward drift of the electron cloud generated by interactions between protons and the gas. A cathode at the bottom of the gas-volume attracted the gas ions. The anode signals were proportional to the energy loss $dE$ of the proton. Above the anodes were PC boards with position-pads connected over delay lines with a 5 ns delay between each pad, which produced a time signal proportional to position in the dispersive direction along the focal plane, proportional to the proton momentum. A planar plastic scintillator detector measured the total energy $E$ deposited by particles passing through the ion chamber. Protons were separated from deuterons reaching the focal plane detector by cuts on the $E$ vs. $dE$ spectrum.

The only detectable contaminant in the target was $^{48}$Ti, which is the titanium isotope with the greatest natural abundance.  A measurement of the
$^{48}$Ti($d,p$)$^{49}$Ti reaction was performed with an enriched target at the same beam energy (16 MeV) to allow the identification of 
contaminant peaks in the $^{50}$Ti($d,p$)$^{51}$Ti spectrum.  

\section{Experimental results}

\begin{figure*}[th]
  \begin{center}
    \scalebox{1.2}{
      \includegraphics{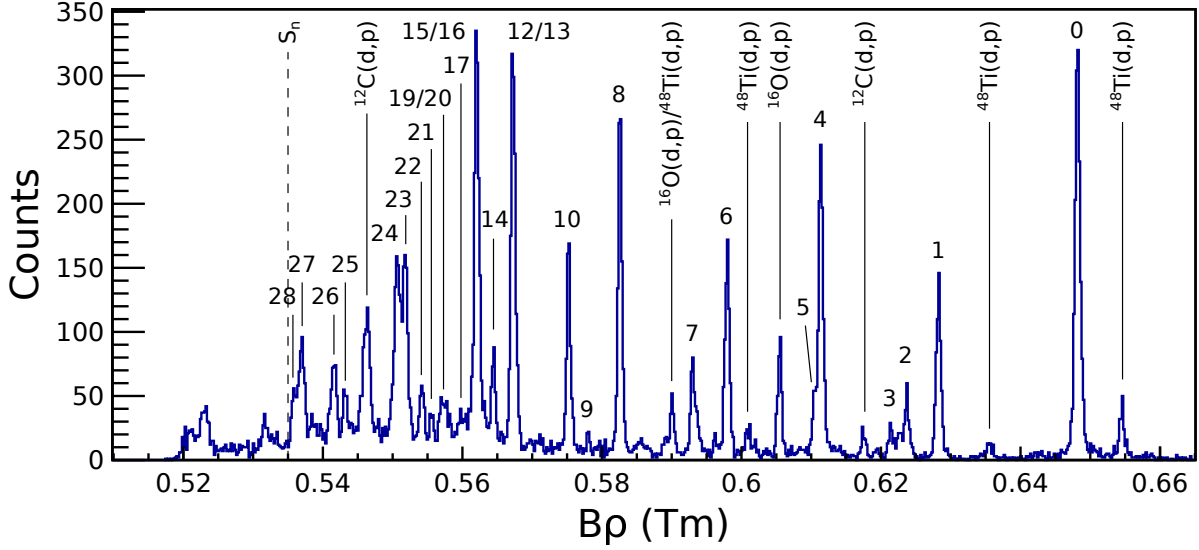}
    }
    \caption{\label{fig:spectrum} (Color online) Proton momentum spectrum at the laboratory angle of 25$^\circ$. Peaks corresponding to states of $^{51}$Ti are labeled from 0 to 28. $B\rho$ is the magnetic rigidity.} 
  \end{center}
\end{figure*}

A representative proton magnetic rigidity spectrum collected at a scattering angle of 25$^\circ$ is shown in Figure~\ref{fig:spectrum}. Peaks are labeled according to the scheme used in Table~\ref{tab:states}. We have adopted the labels used in Ref.~\cite{Bar64} for states 0-21. States 22-28 have been observed via the $(d,p)$ reaction for the first time in the present work. We used two-body kinematics and the energies of states in $^{51}$Ti and $^{13}$C with energies known to $<$1~keV precision to perform a magnetic rigidity calibration to determine proton momentum.  The unlabeled peaks in Figure~\ref{fig:spectrum} are from the $^{48}$Ti contaminant.

\begin{table}[h!]
  \centering
  \caption{\label{tab:states} Excitation energies, angular momentum and $J^\pi$ assignments, single-neutron orbits used for the FRESCO analysis and the spectroscopic factors for states of \textsuperscript{51}Ti populated in the present work. Established $J^\pi$ assignments are from Ref.~\cite{NNDC51}. Tentative $J^\pi$ assignments based on $L$ values determined in the present work are discussed in the text. When more than one possible orbit is given for a state, the spectroscopic factors assuming both orbits are shown.}
  \begin{tabular}{cccccccc}
    \hline\hline
    Label & $E_x$ (keV) & $E_x$(keV) & \textit{L} & J$^\pi$ & orbit & \multicolumn{2}{c}{$S$} \\
          &  (present work) &  (Ref. \cite{NNDC51}) &      &       & & DWBA & ADW \\ \hline
        0  & 0           & 0              & 1 & $\frac{3}{2}^-$ & $p_{3/2}$ & 0.48(6) & 0.47(6) \\
        1  & 1169(3)  & 1166.7(3) & 1 & $\frac{1}{2}^-$ & $p_{1/2}$ & 0.32(4) & 0.32(4) \\
        2  & 1429(4)  & 1437.3(3) &   & $\frac{7}{2}^-$ &  & &         \\
        3  & 1569(4)  & 1567.5(3) &   & $\frac{5}{2}^-$ &  & &       \\
        4  & 2143(2)  & 2144.0(3) & 3 & $\frac{5}{2}^-$ & $f_{5/2}$ & 0.22(3)  & 0.21(3)  \\
        5  & 2197(3)  & 2198.1(4) & 1 & $\frac{3}{2}^-$ & $p_{3/2}$ & 0.039(5) & 0.030(5) \\
        6  & 2908(2)  & 2905.8(5) & 1 & $\frac{1}{2}^-$ & $p_{1/2}$ & 0.22(3)  & 0.19(3)  \\
        7  & 3171(3)  & 3173.8(5) & 1 & $\frac{3}{2}^-$ & $p_{3/2}$ & 0.071(9) & 0.053(7) \\
        8  & 3760(3)  & 3771.3(6) & 4 & $\frac{9}{2}^+$ & $g_{9/2}$ & 0.20(3)  & 0.18(3)  \\
        9  & 4016(4)  & 4022(10)  & 2 & ($\frac{5}{2}^+$) & $d_{5/2}$ & 0.0026(3) & 0.0023(3) \\
        10 & 4166(3)  & 4172(10) & 2 & ($\frac{5}{2}^+$,$\frac{3}{2}^+$) & $d_{5/2}$ & 0.031(4) & 0.029(4) \\
           &          &          &   &                                   & $d_{3/2}$ & 0.049(6) & 0.042(5) \\
        12 & 4567(6)  & 4569(10) & 1 & ($\frac{3}{2}^-$,$\frac{1}{2}^-$) & $p_{3/2}$ & 0.021(3) & 0.015(2) \\
           &          &          &   &                                   & $p_{1/2}$ & 0.040(5) & 0.031(4) \\
        13 & 4602(4)  & 4602(10) & 2 & ($\frac{5}{2}^+$,$\frac{3}{2}^+$) & $d_{5/2}$ & 0.061(8) & 0.055(7) \\
           &          &          &   &                                   & $d_{3/2}$ & 0.094(5) & 0.080(4) \\
        14 & 4751(4)  & 4757(10) &    & &  & & \\
        15 & 4820(5)  & 4820(10) & 3 & ($\frac{5}{2}^-$) & $f_{5/2}$ & 0.024(3) & 0.020(3) \\
        16 & 4882(4)  & 4882(10) & 3 & ($\frac{5}{2}^-$) & $f_{5/2}$ & 0.18(2)  & 0.15(2) \\
        17 & 5001(5)  & 4998(10) & 3 & ($\frac{5}{2}^-$) & $f_{5/2}$ & 0.031(4) & 0.027(3) \\
        19 & 5109(7)  & 5102(10) & 1 & ($\frac{3}{2}^-$,$\frac{1}{2}^-$) & $p_{3/2}$ & 0.019(3) & 0.014(2) \\
           &          &          &   &                                   & $p_{1/2}$ & 0.037(5) & 0.030(4) \\
        20 & 5154(5)  & 5149(10) &  & & & & \\
        21 & 5231(5)  & 5224(10) & 1 & ($\frac{3}{2}^-$,$\frac{1}{2}^-$) & $p_{3/2}$ & 0.014(2) & 0.010(1) \\
        &          &          &   &                                      & $p_{1/2}$ & 0.028(4) & 0.021(3) \\
        22 & 5303(6)  &               & 4 & ($\frac{9}{2}^+$) & $g_{9/2}$ & 0.021(3)  & 0.018(3) \\
        23 & 5427(6)  &               & 3 & ($\frac{5}{2}^-$) & $f_{5/2}$ & 0.078(10) & 0.065(8) \\
        24 & 5492(6)  &               &    &  & & & \\
        25 & 5879(8)  &               &    &  & & & \\
        26 & 5968(7)  &               &   &  &  & & \\
        27 & 6206(8)  &               &  & & & & \\
        28 & 6260(10) &               &  & & & & \\
        \hline\hline
    \end{tabular}
\end{table}

\begin{figure*}[h]
  \begin{center}
    \scalebox{0.6}{
      \includegraphics{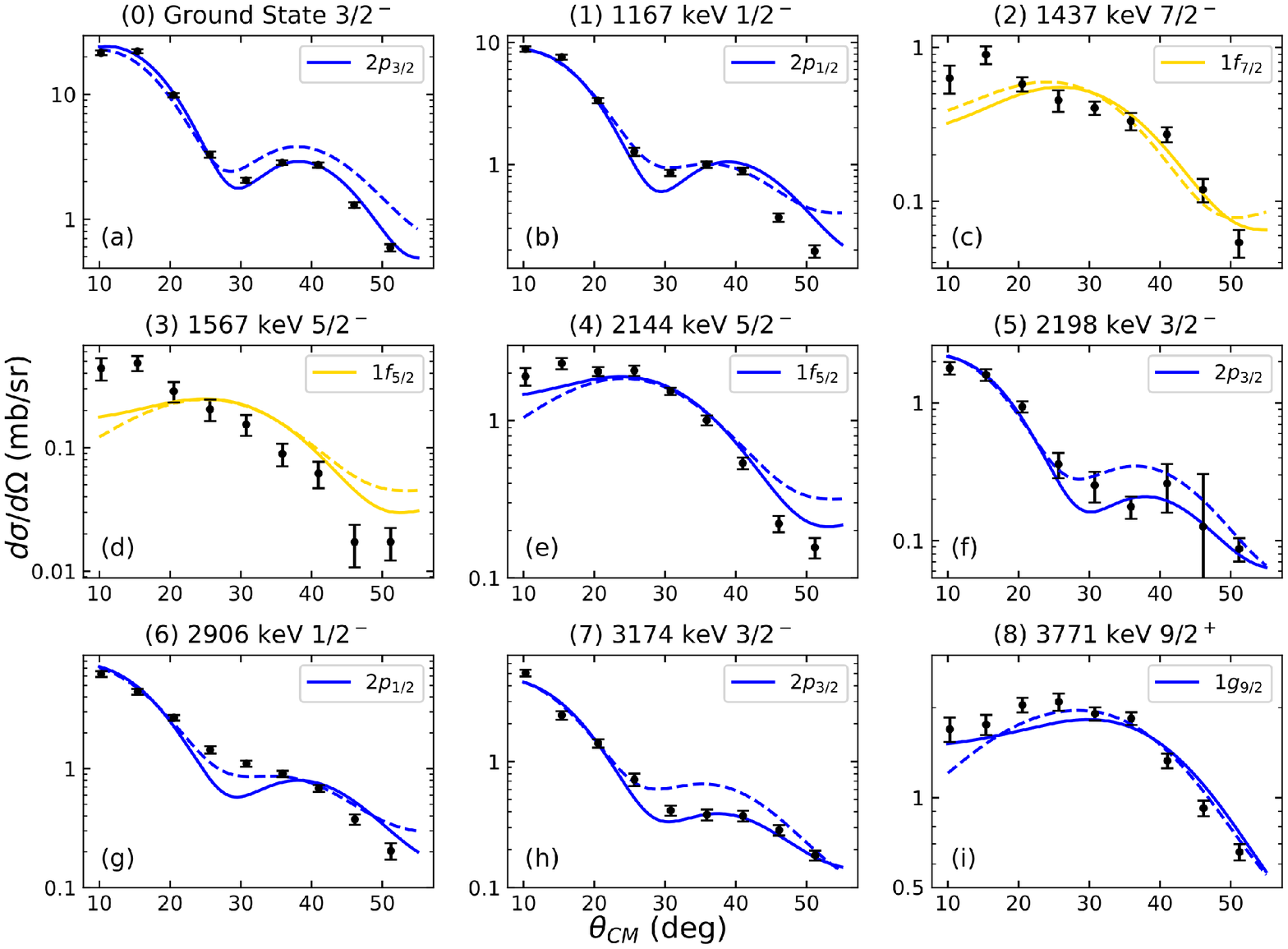}
    }
    \caption{\label{fig:angdist1} (Color online) Measured proton angular distributions from the $^{50}$Ti$(d,p)^{51}$Ti reaction compared with FRESCO calculations described in the text. Panels (a)-(i) correspond to states 0-8 in Table~\ref{tab:states}. The solid curves are ADW calculations, and the dashed curves are DWBA calculations. The excitation energies shown are taken from Ref. \cite{NNDC51}.}
    \end{center}
\end{figure*}

\begin{figure*}[h]
  \begin{center}
    \scalebox{0.6}{
      \includegraphics{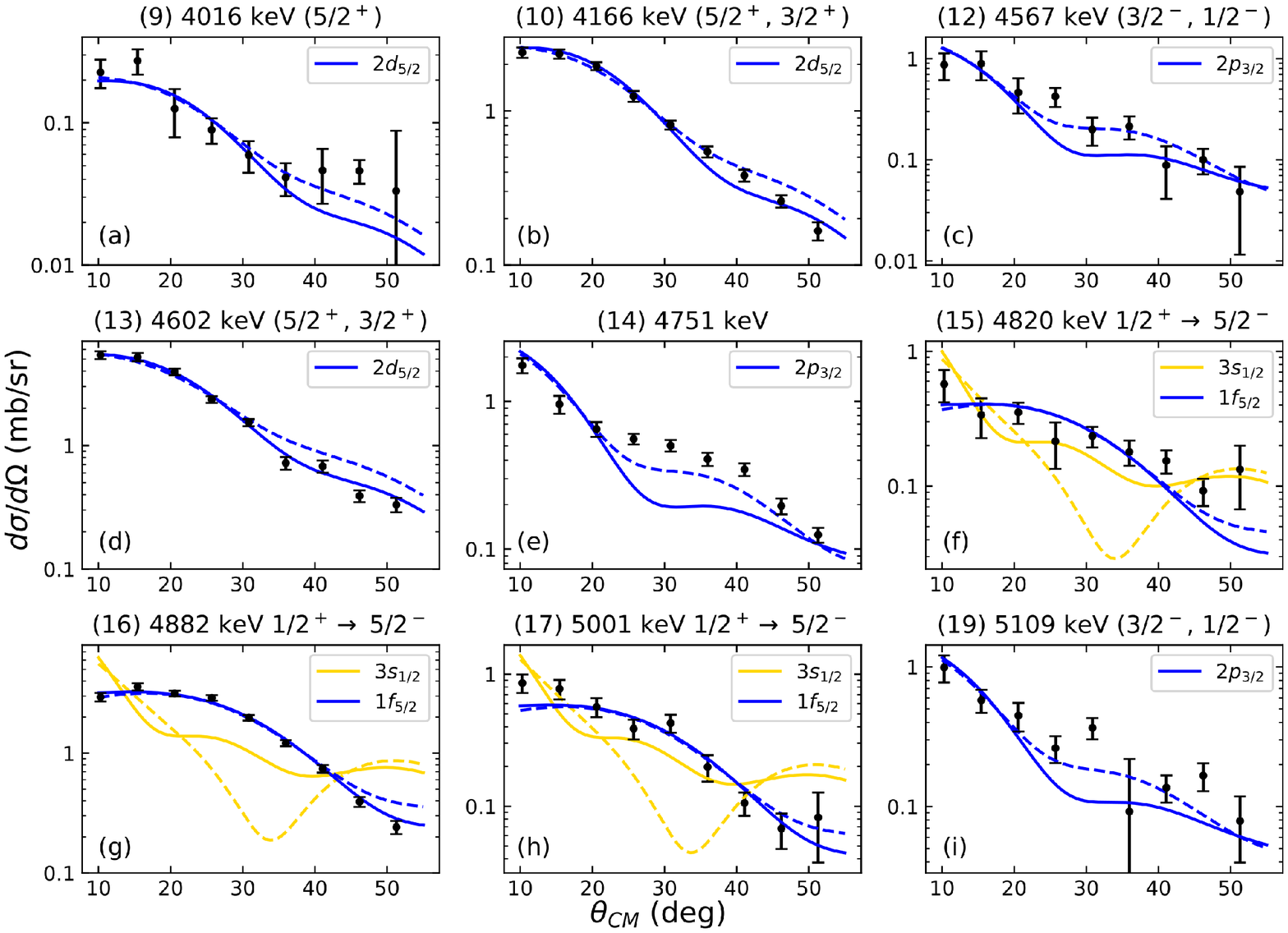}
    }
    \caption{\label{fig:angdist2} (Color online) Measured proton angular distributions from the $^{50}$Ti$(d,p)^{51}$Ti reaction compared with FRESCO calculations described in the text. Panels (a)-(i) correspond to states 9-19 in Table~\ref{tab:states}. The solid curves are ADW calculations, and the dashed curves are DWBA calculations. The excitation energies shown are from the present work.}
    \end{center}
\end{figure*}

\begin{figure*}[h]
  \begin{center}
    \scalebox{0.6}{
      \includegraphics{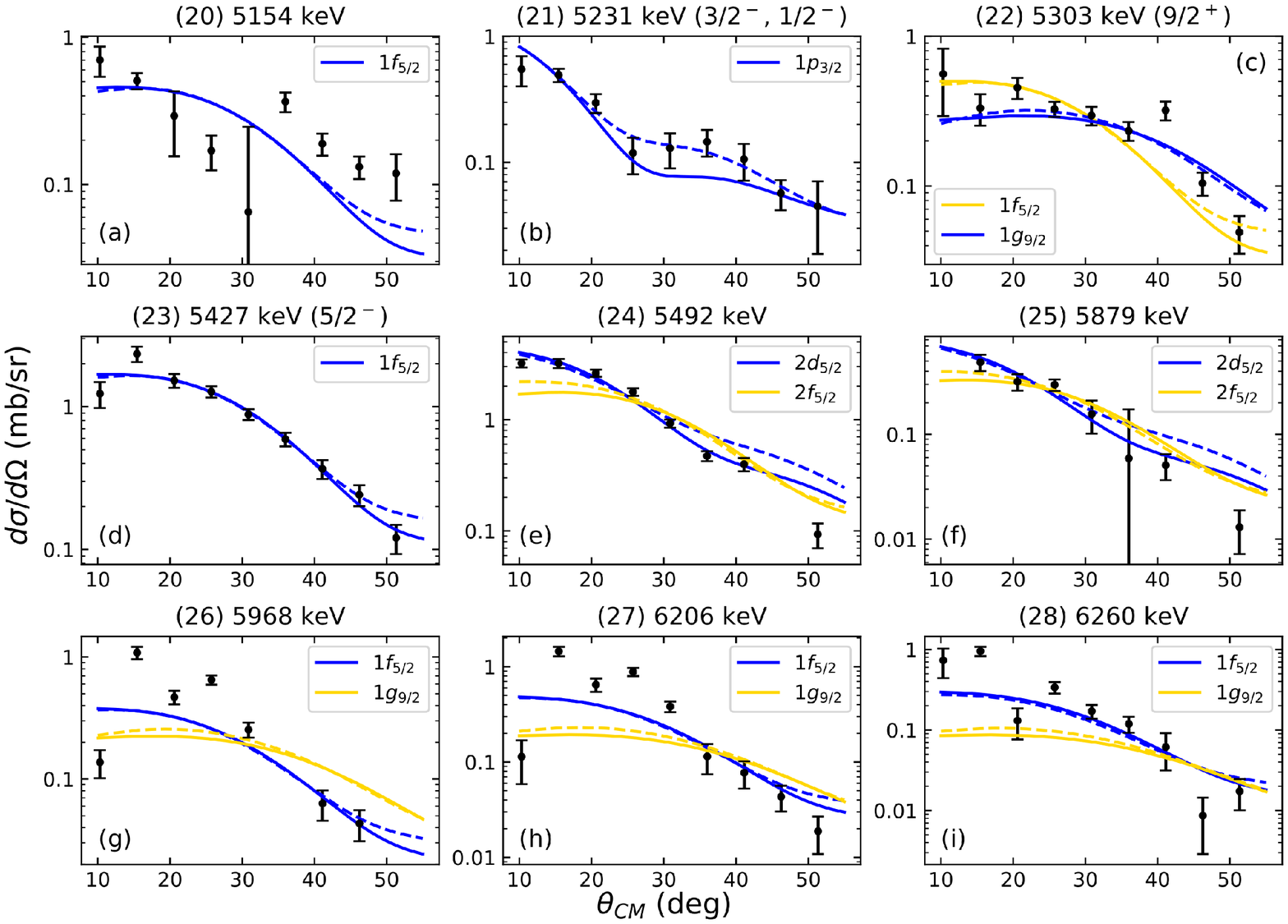}
    }
    \caption{\label{fig:angdist3} (Color online) Measured proton angular distributions from the $^{50}$Ti$(d,p)^{51}$Ti reaction compared with FRESCO calculations described in the text. Panels (a)-(i) correspond to states 20-28 in Table~\ref{tab:states}. The solid curves are ADW calculations, and the dashed curves are DWBA calculations.  The excitation energies shown are from the present work.}
    \end{center}
\end{figure*}

\begin{table*}
  \caption{\label{tab:omps} Optical potential parameters used in FRESCO calculations in the present work.}
  \begin{tabular}{ccccccccccccccc}\hline\hline
             & $V_V$ & $r_V$ & $a_V$& $W_V$ & $r_W$ & $a_W$ & $W_D$ & $r_D$ & $a_D$ & $V_{so}$ & $W_{so}$ & $r_{so}$ & $a_{so}$ & $r_C$  \\
             & (MeV) & (fm) & (fm)  & (MeV) & (fm)  & (fm)  & (MeV)& (fm)  & (fm)  & (MeV)   & (MeV)    & (fm)     & (fm)    & (fm)   \\\hline
    ADW (d) & 104.2 & 1.19 & 0.702 & 1.24  & 1.19  & 0.702 & 15.0 & 1.28  & 0.582 & 11.3    & -0.013   & 1.01     & 0.621   & 1.27   \\
    DWBA (d) &  89.6 & 1.17 & 0.736 & 0.319 & 1.32  & 0.748 & 12.3 & 1.32  & 0.748 & 6.87    &          & 1.07     & 0.660   & 1.30   \\
    DWBA (p) &  53.9 & 1.19 & 0.670 & 1.30  & 1.19  & 0.670 & 8.65 & 1.28  & 0.544 & 5.52    & -0.067   & 1.01     & 0.590   & 1.27   \\\hline\hline
  \end{tabular}
\end{table*}

The magnetic rigidity spectrum measured at each scattering angle was fit using a linear combination of Gaussian functions with a quadratic background. The proton yields corresponding to each state in \textsuperscript{51}Ti were used to produce the measured proton angular distributions shown in Figures~\ref{fig:angdist1}-\ref{fig:angdist3}.  The absolute cross sections were determined to be accurate to an uncertainty of $13\%$, with contributions from uncertainties in charge integration, target thickness and solid angle. 

To extract spectroscopic factors from the present angular distributions, calculations that use the adiabatic approach for generating the entrance channel deuteron optical potentials (as developed by Johnson and Soper \cite{Joh70}) were used.  The potential was produced using the formulation of Wales and Johnson \cite{Wal76}.  Its use takes into account the possibility of deuteron breakup and has been shown to provide a more consistent analysis as a function of bombarding energy \cite{De05} as well as across a large number of ($d,p$) and ($p,d$) transfer reactions on $Z=3-24$ target nuclei \cite{Le07}.  The proton-neutron and neutron-nucleus global optical potential parameters of Koning and Delaroche \cite{Kon03} were used to produce the deuteron potential as well as the proton-nucleus optical potential parameters needed for the exit channel of the ($d,p$) transfer calculations.  In keeping with the nomenclature of Ref. \cite{De05}, these calculations are called ADW.  The angular momentum transfer and spectroscopic factors found in Table ~\ref{tab:states} were determined by fitting these ADW calculations, made with the FRESCO code \cite{Tho88}, to the proton angular distributions.  Since several of the transfer calculations result in $L$ transfers different from those previously reported, standard Distorted Wave Born Approximation (DWBA) calculations were carried out with deuteron entrance channel parameters from Ref. \cite{Dae80} and the same exit proton potentials as in the ADW calculations.  The ADW descriptions of the angular distributions were generally superior at larger angles but the extracted spectroscopic factors were within 20\% of each other.  Optical potential parameters are listed in Table~\ref{tab:omps}.  The overlaps between 
$^{51}$Ti and $^{50}$Ti$+n$ were calculated using binding potentials of Woods-Saxon form whose depth was varied to reproduce the given state's binding energy with geometry parameters of $r_0=1.25$ fm and $a_0=0.65$ fm and a Thomas spin-orbit term of strength $V_{so}=6$ MeV that was not varied.

The present study was motivated in part by our interest in the strong state that Barnes \textit{et al.} \cite{Bar64} reported at 5139 keV.  We did not observe any strong states at or near this energy, although we observed a weak state at 5154(5) keV.  We were unable to make an $L$ assignment for this state.

The $L$ assignments we make here differ from those of Barnes \textit{et al.} for four states.  Barnes \textit{et al.} made a tentative $L=3$ assignment for the state they reported at 4012 keV.  We observed a state at 4016(4) keV and made an assignment of $L=2$ because this value fitted the forward angle data points better.  Barnes \textit{et al.} gave $L=0$ assignments for states they observed at 4810 and 4872 keV.  We observed those states as well (at 4820(5) and 4882(4) keV), but $L=0$ clearly does not fit the measured angular distributions for these states.  Instead, we have made $L=3$ assignments for these states.  Barnes \textit{et al.} also made a tentative $L=0$ assignment for a state at 4988 keV.  We made an $L=3$ assignment for this state instead (which we measured to occur at 5001(5) keV).  Finally, Barnes \textit{et al.} were unable to make an $L$ assignment to the state they observed at 5214 keV.  We measured this state at 5231(5) keV and determined it to have $L=1$.

We found seven new states and were able to make $L$ assignments for two of them.  We made an $L=4$ assignment for the new state at 5303(6) keV and an $L=3$ assignment for the new state at 5427(6) keV.

\section{Single neutron energies in $^{51}$Ti}

The single particle strength for a particular neutron orbital is generally fragmented among several states.  
The ($d$,$p$) reaction reveals the states in which those fragments are located and allows the determination of spectroscopic factors 
for those states so that the single neutron energy for an orbital can be calculated as the centroid of the fragments.
  
The largest concentration of $p_{3/2}$ neutron transfer strength in $^{51}$Ti is located in the ground state.  
However, fragments of the $p_{3/2}$ strength are located in the 2198 and 3174 keV states. (In this discussion and the calculations of
single neutron strength centroids, we use the spectroscopic factors from the ADW analysis.  We also use the energies from Ref. \cite{NNDC51} for states labeled 0-8, and the energies from the present work for all others.)  Furthermore, 
there is $L=1$ transfer strength in the 4567, 5109 and 5231 keV states.  However, a ($d$,$p$) study with a 
polarized deuteron beam would be required to determine whether the $L=1$ strength in these states comes 
from the $p_{3/2}$ or $p_{1/2}$ orbitals.  This introduces an uncertainty into the result for 
the $p_{3/2}$ orbital:  The lowest possible centroid for the $p_{3/2}$ state is given by assuming that the 4567, 5109 and 5231 keV states 
are $p_{1/2}$, which gives 423 keV above the ground state energy.  The highest possible centroid for the $p_{3/2}$ state is calculated assuming 
that all three of these high-lying states are $p_{3/2}$, giving 720 keV above the ground state energy.  Therefore, the centroid of 
the $p_{3/2}$ strength is given by an energy of 572(149) keV above the ground state.  Since the ground state 
has a binding energy of 6372 keV, we find that the binding energy of the $p_{3/2}$ neutron orbital is 5800(149) keV.  

The largest concentration of $p_{1/2}$ strength is found in the first excited state at 1167 keV, but the 2906 keV state 
has a significant amount of $p_{1/2}$ strength as well.  If these two states are the only states with $p_{1/2}$ strength, then the centroid is 1815 keV above the ground state.  If the 4567, 5109 and 5231 keV states have $J^{\pi}=1/2^-$, then the centroid is 2247 keV.  Therefore, 
our result for the $p_{1/2}$ single neutron energy is 2031(216) keV above the ground state, 
corresponding to a binding energy of 4341(216) keV.  

The 2144 keV state is the only state with an $L=3$ transfer that has a definitive $5/2^-$ assignment, 
and it has a spectroscopic factor of 0.21(3) for $f_{5/2}$ transfer.  However, there are 
four $L=3$ states at 4820, 4882, 5001 and 5427 keV for which definitive spin assignments are not available.  
It is likely that they are $f_{5/2}$ states because the $f_{7/2}$ neutron orbital is located below 
the $N=28$ shell closure and is therefore presumed to be full.  There are two $7/2^-$ states 
in $^{51}$Ti at relatively low energies (1437 and 2691 keV).  However, both states are seen strongly 
in the $^{49}$Ti($t$,$p$) reaction \cite{Gl68}, which provides convincing evidence that their structure can be 
described as a pair of $p_{3/2}$ neutrons coupled to spin zero coupled to an $f_{7/2}$ neutron hole.
Furthermore, these states are only weakly populated in the present ($d$,$p$) experiment.  
If we assume that the 4820, 4882, 5001 and 5427 keV states have $J^{\pi}=5/2^-$, the centroid of 
the $f_{5/2}$ neutron strength (and therefore the single neutron energy) is 3743 keV above 
the ground state, corresponding to a binding energy of 2629 keV.  For the remainder of the discussion in this paper, we will assume that this is the case.  A measurement of the $^{49}$Ti($t,p$)$^{51}$Ti reaction could, in principle, conclusively determine whether these states have $J^{\pi}=5/2^-$ or $7/2^-$.  Unfortunately, the study reported in \cite{Gl68} only observed states up to 3.0 MeV.  It is clear that this reaction should be revisited, this time with access to higher excitation energies.      

The $N=32$ subshell gap is the gap between the $p_{3/2}$ orbit and the next higher orbit, which in $^{51}$Ti is the 
$p_{1/2}$ orbit.  To determine what this gap is, we must consider the uncertainty in the $J^{\pi}$ assignments for the 4567, 5109 and 5231 keV states.  The gap is smallest if all three of these states have $J^{\pi}=3/2$.  In that case, the centroid for the $p_{3/2}$ orbit is 720 keV above the ground state and the centroid for the $p_{1/2}$ orbit is 1815 keV above the ground state, giving a subshell gap of 1095 keV.  The largest possible gap, calculated assuming that the 4567, 5109 and 5231 keV states have $J^{\pi}=1/2^-$, is 1824 keV.  Therefore, our result for the size of the $N=32$ subshell gap is 1459(365) keV.  

Assuming that the 4820, 4882, 5001 and 5431 keV states have $J^{\pi}=5/2^-$ (so that the centroid for $f_{5/2}$ is 3743 keV above the ground state), the $f_{5/2}$ orbit is 3171(149) keV above the $p_{3/2}$ orbit.

We close this section with the caveat that it is possible that we have not observed all the weak fragments of the neutron orbits we have examined here.  The possibility that such fragments exist - particularly at the higher energies studied here - introduces a further source of uncertainty to our results.

\section{Discussion}

\begin{figure}[h]
  \begin{center}
    \scalebox{0.33}{
      \includegraphics{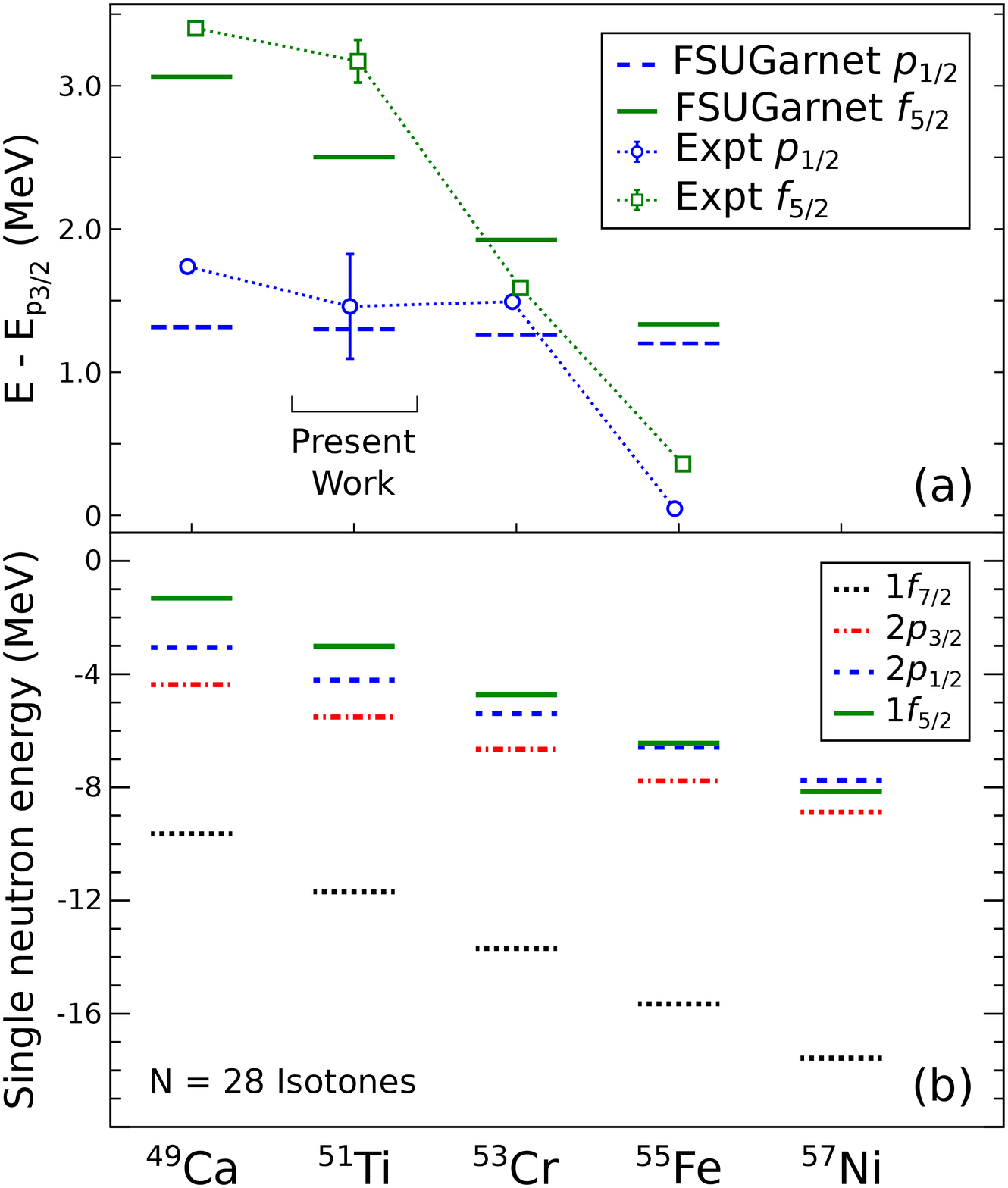}
    }
    \caption{\label{fig:sn_energies} (Color online) (a) Measured $f_{5/2}$ and $p_{1/2}$ single neutron energy centroids, relative to the $p_{3/2}$ energy, from the present work and Refs.~\cite{NNDC49,NNDC53,NNDC55} compared with the covariant density functional theory approach described in the text. (b)  Single neutron binding energies calculated using the covariant density functional theory.} 
    \end{center}
\end{figure}

Single particle energies are not static.  Instead, they vary as a function of proton and neutron numbers.  
The top panel of Figure~\ref{fig:sn_energies} shows the present results for $^{51}$Ti and the single
neutron energy centroids for the $p_{1/2}$ and $f_{5/2}$ orbitals relative to the $p_{3/2}$ orbital 
for the $N=29$ isotones $^{49}$Ca, $^{53}$Cr and $^{55}$Fe from ($d$,$p$) results compiled 
by the National Nuclear Data Center \cite{NNDC49,NNDC53,NNDC55}. 

These experimental results require some explanation.  The $^{48}$Ca($d,p$)$^{49}$Ca measurement \cite{Uo94} cited in Ref. \cite{NNDC49} was performed with polarized deuterons, so that there is no uncertainty in the spins of the measured states.  

The situation in $^{53}$Cr is similar to that in $^{51}$Ti:  There are states with $L=1$ and $L=3$ for which the spins are uncertain.  That is, some of the $L=1$ states (at 2454, 2723, 3587, 4610 and 5557 keV) might have $J^{\pi}$ values of either $1/2^-$ or $3/2^-$.  The gap between the $p_{3/2}$ and $p_{1/2}$ single neutron energies is a minimum if all five of these states have $J^{\pi}=3/2^-$, and that minimum gap is 1162 keV.  The gap is a maximum if all five of those states have $J^{\pi}=1/2^-$ - that maximum gap is 1813 keV.  Therefore, the $p_{3/2}-p_{1/2}$ gap in $^{53}$Cr is 1488(326) keV.  As we did in $^{51}$Ti, we assume that all of the $L=3$ states with unknown $J^{\pi}$ values (at 2664, 3005 and 4666 keV) have $J^{\pi}=5/2^-$.  That gives a $p_{3/2}-f_{5/2}$ energy difference of 1424(165) keV - a much smaller energy difference than in $^{51}$Ti, where that difference is 3138(184) keV.

The $^{54}$Fe($d,p$)$^{55}$Fe reaction has been studied with a polarized beam, so there is considerably more certainty regarding $J^{\pi}$ values.  There are spectroscopic factors determined for five $3/2^-$ states (the ground state and excited states at 2052, 2471, 3035 and 3553 keV), which give a $p_{3/2}$ centroid of 581 keV above the ground state.  Spectroscopic factors have been determined for three $1/2^-$ states (413, 1919 and 5775 keV), which give a centroid for the $p_{1/2}$ neutron orbit of 939 keV.  This results in a gap of only 358 keV between these spin-orbit partners, which is much smaller than the corresponding gaps in $^{51}$Ti and $^{53}$Cr of about 1.4 MeV.  It is unlikely that the spin-orbit splitting changes this dramatically so quickly; therefore, a remeasurement of the $^{54}$Fe($d,p$)$^{55}$Fe reaction should be performed.

Spectroscopic factors have been determined for three $5/2^-$ states in $^{55}$Fe (at 933, 2144 and 4057 keV), which gives a centroid for the $f_{5/2}$ neutron orbit of 1360 keV above the ground state - only 779 keV above the $p_{3/2}$ single neutron energy.        

Figure 5a also includes the results of theoretical predictions made in the framework of covariant density functional theory.  
The bottom panel of Figure~\ref{fig:sn_energies} shows the theoretical predictions as binding energies.  In covariant density functional
theory, the basic constituents are protons and neutrons interacting via the 
exchange of various self-interacting mesons and the photon. Nucleons satisfy a Dirac 
equation in the presence of strong scalar and vector potentials that are the hallmark 
of the relativistic approach. In particular, the strong potentials provide a natural explanation 
for the strong spin-orbit splitting characteristic of atomic nuclei. Equally natural within the 
relativistic framework is the explanation of the \emph{pseudo-spin} symmetry that encodes 
the relatively small energy gaps of pseudo-spin-orbit partners 
($s_{1/2}\!-\!d_{3/2}$, $p_{3/2}\!-\!f_{5/2}$, etc.)\,\cite{gin96}. Indeed, whereas spin-orbit 
partners have upper components of Dirac orbitals that share the same value of the orbital 
angular momentum (e.g., $l\!=\!1$ for $p_{3/2}\!-\!p_{1/2}$), it is the orbital angular 
momentum of the \emph{lower components} that is the same for pseudo-spin-orbit 
partners (e.g., $l\!=\!2$ for $p_{3/2}\!-\!f_{5/2}$).

The evolution of single-particle gaps was predicted using the covariant energy density 
functional FSUGarnet\,\cite{che15} that was calibrated using the fitting protocol described 
in Ref.\,\cite{che14}. In a mean-field approximation one must solve a non-linear set of
differential equation self-consistently. That is, the single-particle orbitals satisfying the 
Dirac equation are generated from the various meson fields which, in turn, satisfy 
Klein-Gordon equations with the appropriate ground-state densities as the source terms. 
The outcome of such an iterative procedure are self-consistent potentials, ground-state 
densities, and the binding energies that have been plotted in Figure~\ref{fig:sn_energies}.
We note that the isovector sector of the nuclear energy density functional (namely, the
component that distinguishes protons from neutrons) is poorly constrained. Hence, studies that
examine the evolution of experimental quantities as functions of neutron excess like the ones
carried out here provide important constraints on the isovector sector of nuclear
models.

The calculations provide support for the notion that the spacing between the $p_{3/2}$ and $f_{5/2}$ neutron orbits shrinks dramatically as the proton number increases.  However, the calculations also predict that the splitting between the $p_{3/2}$ and $p_{1/2}$ orbits remains constant along the isotonic chain - reinforcing the idea that the $^{54}$Fe($d,p$)$^{55}$Fe reaction should be remeasured.

\section{Conclusions}

We performed a measurement of the $^{50}$Ti($d$,$p$)$^{51}$Ti reaction at 16 MeV 
using a Super Enge Split Pole Spectrograph.  Seven states were observed that had 
not been observed in previous ($d$,$p$) measurements, and the \textit{L} transfer values 
for six previously measured states were either changed or measured for the first time.  The results provide support for the existence of the $N=32$ subshell gap in Ti isotopes.  However, a measurement of the $^{49}$Ti($t,p$)$^{51}$Ti reaction above 3 MeV excitation energy would allow a more precise measurement of the $f_{5/2}$ single neutron energy by distinguishing between $J^{\pi}=5/2^-$ and $7/2^-$ values for the high-lying $L=3$ states observed in the present ($d,p$) reaction.  Furthermore, previous measurements of the $^{54}$Fe($d,p$)$^{55}$Fe reaction show a collapse in the gap 
between the $p_{3/2}$ and $p_{1/2}$ spin-orbit partners in $^{55}$Fe, which is not predicted in a covariant density functional calculation.  The 
$^{54}$Fe($d,p$)$^{55}$Fe reaction should be revisited.

\begin{acknowledgments}
A target provided by the Center for Accelerator Target Science at Argonne National Laboratory was used in this work.  We thank Jorge Piekarewicz for the covariant density functional theory calculations.  We also thank Professor Nicholas Keeley for productive discussions of the ADW calculations and continuing advice.  This work was supported by the National Science Foundation through grant numbers PHY-1712953 and PHY-1617250.
\end{acknowledgments}


%

\end{document}